\documentclass[sigconf]{acmart}
\usepackage{multirow}
\usepackage{algorithm}
\usepackage{algorithmic}

\AtBeginDocument{%
  \providecommand\BibTeX{{%
    \normalfont B\kern-0.5em{\scshape i\kern-0.25em b}\kern-0.8em\TeX}}}

\copyrightyear{2021}
\acmYear{2021}
\setcopyright{acmcopyright}
\acmConference[SIGIR '21] {Proceedings of the 44th International ACM SIGIR Conference on Research and Development in Information Retrieval}{July 11--15, 2021}{Virtual Event, Canada.}
\acmBooktitle{Proceedings of the 44th International ACM SIGIR Conference on Research and Development in Information Retrieval (SIGIR '21), July 11--15, 2021, Virtual Event, Canada}
\acmPrice{15.00}
\acmDOI{10.1145/3404835.3462979}
\acmISBN{978-1-4503-8037-9/21/07}

\begin{document}
\fancyhead{}

\title{Towards a Better Tradeoff between Effectiveness and Efficiency in Pre-Ranking: A Learnable Feature Selection based Approach}


 \author{
	Xu Ma$^{\dagger}$, Pengjie Wang$^{\dagger}$, Hui Zhao, Shaoguo Liu, Chuhan Zhao, Wei Lin, Kuang-Chih Lee,\\ Jian Xu$^{\ddagger}$ and Bo Zheng$^{*}$
 }
 \thanks{$\dagger$  Joint first authors.}
 \thanks{ $\ddagger$ This author gave a lot of guidance in the work.}
 \thanks{$*$ Corresponding author}
 \thanks{Project funded by China Postdoctoral Science Foundation (2019TQ0290)}
 
\affiliation{%
   \institution{Alibaba Group}
 }
 \email{
 {maxu.mx, pengjie.wpj, shuqian.zh, shaoguo.lsg, hz_chuhan.zch, kuang-chih.lee, xiyu.xj, bozheng}@alibaba-inc.com, lwsaviola@163.com
 }

\renewcommand{\shortauthors}{Ma and Wang, et al.}
\renewcommand{\algorithmicrequire}{ \textbf{Input:}}     
\renewcommand{\algorithmicensure}{ \textbf{Output:}}    

\begin{abstract}

In real-world search, recommendation, and advertising systems, the multi-stage ranking architecture is commonly adopted. Such architecture usually consists of matching, pre-ranking, ranking, and re-ranking stages. In the pre-ranking stage, vector-product based models with representation-focused architecture are commonly adopted to account for system efficiency. However, it brings a significant loss to the effectiveness of the system. In this paper, a novel pre-ranking approach is proposed which supports complicated models with interaction-focused architecture. It achieves a better tradeoff between effectiveness and efficiency by utilizing the proposed learnable Feature Selection method based on feature Complexity and variational Dropout (FSCD).
Evaluations in a real-world e-commerce sponsored search system for a search engine demonstrate that utilizing the proposed pre-ranking, the effectiveness of the system is significantly improved. Moreover,  compared to the systems with conventional pre-ranking models, an identical amount of computational resource is consumed.

\end{abstract}

\ccsdesc[500]{Information systems~Learning to rank}

\keywords{pre-ranking, effectiveness, efficiency, feature selection}


\maketitle

\vspace{-2ex}

\section{Introduction}

\begin{figure}[]
\centering
\includegraphics[width=0.5\textwidth, keepaspectratio]{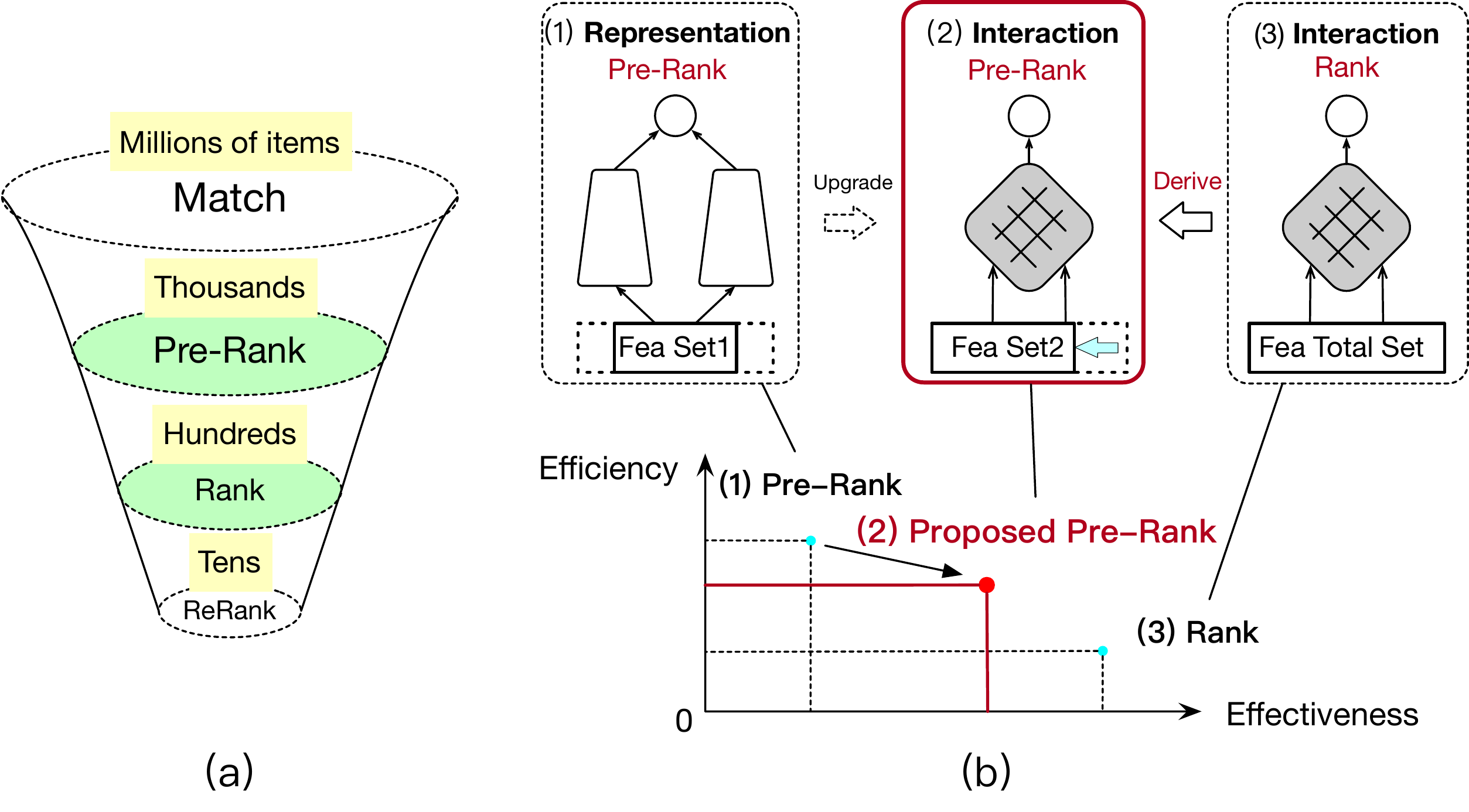}
\vspace{-7ex}
\caption{Real-world multi-stage ranking architecture. (a) Multi-stage architecture including matching, pre-ranking, ranking, and re-ranking with scoring item numbers. (b) Intuitive view for effectiveness and efficiency of pre-ranking and ranking, where the pre-ranking is derived from ranking and optimized to an interaction-focused architecture.}
\label{fig:EE}
\vspace{-9ex}
\end{figure}

As important internet services, large-scale search engine and recommendation systems play important roles in information retrieval and item recommendation, where a ranking system selects only a few items from tens of millions of candidates. Under the constraint of extremely low system  latency, a single complicated ranking model cannot rank the entire candidate set. Therefore, multi-stage ranking architecture is commonly adopted~\cite{raykar2010designing,liu2017cascade,gallagher2019joint}
(shown in Figure~\ref{fig:EE} (a)). Large-scale deep neural networks (DNNs) with interaction-focused architecture~\cite{guo2020deep} are usually employed for the ranking model to maintain good system performance, while only less complicated models with representation-focused architecture~\cite{guo2020deep} are adopted in pre-ranking to ensure efficiency. 

However, simple pre-ranking models with representation-focused architecture will inevitably diminish the model expression ability. Vector-product based model~\cite{huang2013learning,wu2018eenmf}, which is classified as representation-focused architecture, is often employed in pre-ranking (see Figure~\ref{fig:EE} (b)).  The limitation is that neither explicit interactive features nor implicit interactive semantics, which are less efficient for computation but very effective for the expression ability~\cite{wang2018billion}, can be used. 
The pre-ranking models with representation-focused architecture focus excessively on the efficiency optimization and remain a large effectiveness gap to models with interaction-focused architecture. Moreover, as the online feature generation process consume as much resource as that needed for the online inference of the model, the utilization for the pre-ranking with interaction-focused architecture is possible by using feature selection considering both effectiveness and efficiency.
In this context, the pre-ranking with an interaction-focused architecture is studied in this paper.

 
\vspace{-0ex}

In this paper, a pre-ranking model with tradeoff of both effectiveness and efficiency is proposed for a real-world e-commerce application, where the effectiveness improves significantly with slight decrease of efficiency shown in Figure~\ref{fig:EE} (b). The contributions of this paper are summarized as follows. 1) A pre-ranking model with interaction-focused architecture is proposed by inheriting the architecture of ranking model to solve the problem of performance loss of the model with representation-focused architecture. 2) A learnable Feature Selection method based on feature Complexity and variational Dropout (FSCD) is proposed to search for a set of effective and efficient feature fields for the pre-ranking model. 3) Extensive experiments are carried out which show the proposed approach achieves great improvement in effectiveness of the system compared to the conventional baselines, while the efficiency of the system is also maintained.
 
\vspace{-4ex}

\section{The Proposed Approach}\label{sec-approach}

\vspace{-1ex}

The proposed pre-ranking is derived from the ranking model. Both models utilize interaction-focused architectures, which shares an identical feature set $S = \{f_1, f_2, ..., f_{M}\}$, where $f_j$ is the $j$-th feature field in $S$, and $M$ is the number of feature fields. In the system, the ranking model employs all feature fields in $S$ to maintain effectiveness, while the pre-ranking model utilizes only a subset to reduce computational complexity and scores more items. In this context, offline processes such as sample generation can be reused for both models, which saves offline computational resources. The interaction-focused architecture of pre-ranking is inherited from that of ranking model, where both explicit interactive features and implicit interactive semantics can be utilized, which reduces the gap in the models' optimization objectives and achieves significant improvement of the model effectiveness for pre-ranking compared to the model with representation-focused architecture.

For introducing interaction-focused architecture into pre-ranking, a great challenge is the model efficiency. The overall efficiency of the pre-ranking model, whose learnable variables can be divided into feature embeddings $\pmb{v}$ and dense weights $\pmb{w}$, is strongly influenced by the feature embeddings $\pmb{v}$. For example, in real-world scenarios, storage for feature embeddings exceeds over $95\%$ of that for the whole model, while feature generation process for embeddings consumes as much resources as that needed for the online inference of the model. Therefore, the feature selection for features with both high effectiveness and efficiency is important for the utilization of pre-ranking with interaction-focused architecture.


\vspace{-3ex}
\subsection{FSCD for Pre-Ranking Model}\label{feature_selection_method}

\begin{figure}[]
\centering
\includegraphics[width=0.5\textwidth, keepaspectratio]{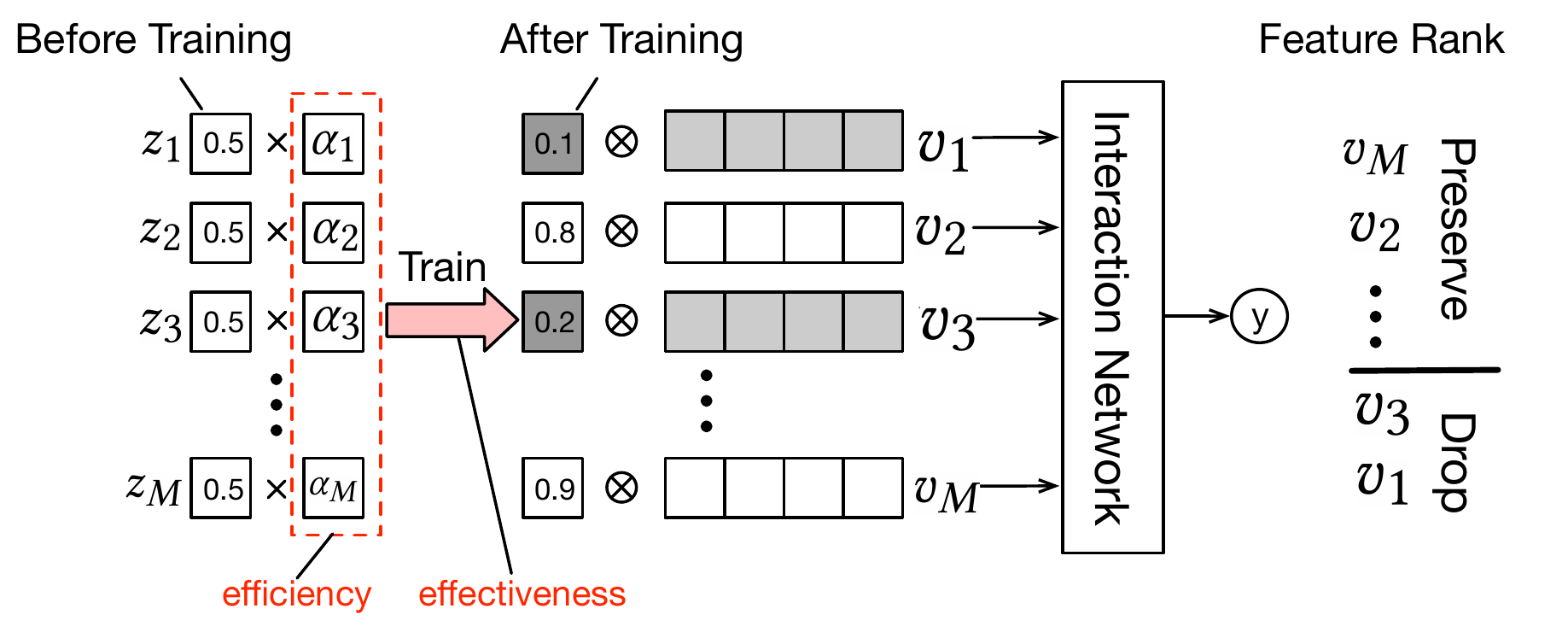}
\vspace{-5ex}
\caption{The proposed FSCD for pre-ranking model.}
\label{fig:fig_structure}
\vspace{-3ex}
\end{figure}

Inspired by the Dropout FR method based on the variational dropout layer~\cite{2017Dropout} where the efficiency of the model is ignored, FSCD is  proposed that considers both effectiveness and efficiency in a learnable process as illustrated in Figure~\ref{fig:fig_structure}. In the proposed FSCD, the effectiveness is optimized by the cross entropy based loss function, while the efficiency is optimized by the feature-wise regularization term in Eq.~(\ref{equ:l_w_z}). Both effective and efficient features can be selected by FSCD in one single training process, while the expression ability of pre-ranking model is improved utilizing these features compared to the vector-product based model. The details are as follows.

To select feature fields with both high effectiveness and high efficiency, each feature field $f_j$ is expected to learn a dropout parameter $z_j \in \{0, 1\}$ to indicate whether the feature field is dropped ($z_j = 0$) or preserved ($z_j=1$). $f_j$'s embeddings $v_j$ are multiplied by the dropout $z_j$ to form the embedding layer, and $z_j$ is subject to a Bernoulli distribution parameterized by $\theta_j$, i.e., 
\begin{equation}\label{Bern}
z_j \sim \textup{Bern}(\theta_j),
\end{equation}
where the hyperparameter $\theta_j$ is the priori probability for the preservation of feature field $f_j$ and is configured as function of feature complexity $c_j$, i.e.,
\begin{equation}\label{theta_j}
\theta_j = \mathcal{H}(c_j) = 1 - \sigma (c_j), \\
c_j = \mathcal{G}(o_j, e_j, n_j)
\end{equation}
where $\sigma(\cdot)$ is the sigmoid function. $\theta_j = 1 - \sigma (c_j)$ is one of alternatives to relate ${\theta_j}$ and ${c_j}$ which works well in practice. The feature complexity $c_j$ measures the computational and storage complexity of the $j$-th feature field including but not limited to the online computational complexity $o_j$, the embedding dimension $e_j$, and the number of keys $n_j$ for one feature field, where $o_j$ is configured according to the feature type specified in Section~\ref{Experiment_configurations}. Linear combination $c_j = \gamma_1 o_j + \gamma_2 e_j + \gamma_3 n_j$ is one of the alternatives with hyperparameter $\gamma_{1,2,3}$. According to Eq.~(\ref{theta_j}), a feature with large complexity has a small value of $\theta_j$, and vice versa. 

\vspace{0ex}

Given the training samples $\mathcal{D}= \{(x_i, y_i) |i = 1,2,...,N \}$, where $N$ is the number of samples, the overall loss function for the learnable feature selection is derived by Bayesian rule~\cite{rendle2012bpr} as,

\vspace{-1ex}
\begin{equation}\label{equ:l_w_z}
\begin{split}
L(\pmb{v}, \!\pmb{w},\! \pmb{z}) \!=\! -\frac{1}{N}\textup{log}P(\mathcal{D}|\pmb{v}, \pmb{w},\pmb{z})  \!+\!  \!\frac{\lambda}{N} \!\!( \!||\pmb{w}||^2 \!\!+\! \! ||\pmb{v}||^2) \!+\!\! \sum_{j=1}^{M} \frac{\alpha_j z_j}{N} \!,
\end{split}
\end{equation}
where $\alpha_j$ is the regularization weight for $z_j$ and can be derived as (See derivation details in Appendix~\ref{appendix})
\begin{equation}\label{alpha_theta}
\alpha_j = \textup{log}(1-\theta_j)- \textup{log}(\theta_j). 
\end{equation}
$\alpha_j$ is a function that decreases with $\theta_j$ and increases with $c_j$. Therefore, a feature with larger complexity $c_j$ is penalized with a larger value of $\alpha_j$, and more likely to be dropped. In this way, the feature complexity is included in the proposed FSCD, which previous works do not address to the best of the authors' knowledge. 

Subject to Bernoulli distribution, $z_j$ is discrete and not differentiable, it is relaxed to a differentiable function as~\cite{jang2016categorical}
\begin{equation}\label{delta}
z_j \!= \!\mathcal{F}(\delta_j) \! = \!\sigma(\!\frac{1}{t}\!(\!\textup{log}(\delta_j)-\textup{log}(\!1-\delta_j\!)+\textup{log}(u_j)-\textup{log}(\!1-u_j\!)\!)\!),
\end{equation}
where $u_j \sim \textup{Uniform}(0,1)$ is subject to a uniform distribution and changes during the training process, while $t = 0.1$ is a constant that works well in the experiments. $\mathcal{F}(\delta_j)$ is close to 0 or 1 for most $\delta_j$ values, which approximates the discrete Bernoulli distribution. In contrast to the dropout $z_j$, $\delta_j \in (0,1)$ is a differentiable parameter. Moreover, it acts as the  posterior probability for feature preservation, which is influenced by the priori probability for feature preservation $\theta_j$, and can be learned as the feature importance.

In this context, the entire process of training for feature selection is built in Eq.~(\ref{equ:l_w_z}) and~(\ref{delta}). Note that, the learnable variables are $\pmb{v}$, $\pmb{w}$ and $\pmb{\delta}$, which are trained simultaneously. Through a fast convergence of $\pmb{\delta}$, the feature set of the pre-ranking model can then be obtained by selecting the feature fields with top-$K$ of $\pmb{\delta}$ values. 

\vspace{-2ex}

\subsection{Fine-Tuning the Pre-Ranking Model}\label{model_training_method}

After feature selection, the feature fields not in the pre-ranking set acquired by FSCD are masked, and the models are fine-tuned using weights $\pmb{v}$ and $\pmb{w}$ as initialization parameters. Concretely, the model 
can be trained with the following loss function
\vspace{-1ex}
\begin{equation}\label{L_s1}
L(\pmb{v}',\pmb{w}') = -\sum_{i=1}^{N}\textup{log}p(y_i|f(x_i', \pmb{v}', \pmb{w}')),
\end{equation}
where $x_i'$, $\pmb{v}'$, and $\pmb{w}'$ are the samples with the selected feature fields, the remaining feature embeddings, and dense weights for the pre-ranking model, respectively. $\pmb{v}'$ and $\pmb{w}'$ are initialized by $\pmb{v}$ and $\pmb{w}$, which accelerates the training. The Bernoulli dropout $\pmb{z}$ is omitted in the fine-tuning process.

In this way, the pre-ranking model is obtained.
As the model is in an interaction-focused architecture and adopts both effective and efficient feature fields, its expression ability significantly improves.
Therefore, the effectiveness of the system can be improved with high efficiency. The entire method is illustrated in Algorithm~\ref{algorithm1}.
\vspace{-0.3cm}
\begin{algorithm}[]
\small
\caption{The Proposed Pre-ranking based on FSCD}\label{algorithm1}
\begin{algorithmic}[1] 
   \REQUIRE Feature set $S = \{f_1, f_2, ..., f_{M}\}$, feature complexity metrics $\{o_j, e_j, n_j\}$, training samples $\mathcal{D}$, hyperparameters $\gamma_{1,2,3}$. 
    \ENSURE Pre-ranking model $\mathcal{M}$.
    \STATE  Calculate $c_j$, $\theta_j$ and $\alpha_j$ for each feature field using Eq.~(\ref{theta_j}) and ~(\ref{alpha_theta}).
    \STATE Build the model with interaction-focused architecture where the feature embeddings are multiplied by $z_j$ parameterized by $\delta_j$ in Eq.~(\ref{delta}).
    \STATE Train using Eq.~(\ref{equ:l_w_z}), and then select the feature fields with top-$K$ of $\delta_j$.
    \STATE Fine-tune according to Eq.~(\ref{L_s1}) and obtain $\mathcal{M}$.
\end{algorithmic}
\end{algorithm}
\vspace{-0.5cm}


\section{Experiments}\label{sec-exp}

\subsection{Experiment Configurations}\label{Experiment_configurations}

Evaluations are mainly based on the online sponsored search system for a real-world e-commerce app named Taobao, where both the click-through rate (CTR) and response per mile (RPM), which evaluates platform revenue, are optimized. The proposed effective and efficient pre-ranking are compared with the baselines of the vector-product based model~\cite{huang2013learning} and COLD~\cite{wang2020cold} based pre-ranking.

Both the pre-ranking and ranking models are based on an interaction-focused architecture. After the feature fields are transformed into embeddings, hidden layers with sizes of $1024$, $512$, and $256$ are adopted for the ranking, while only $2$ hidden layers with sizes of $1024$ and $256$ are used for the pre-ranking to ensure high efficiency. Finally, the sigmoid function is utilized to predict the final CTRs for both models.


The feature set for the pre-ranking model is a subset of that for the ranking model, which is selected by the proposed FSCD. All feature sets, consisting of 246 feature fields in their entirety, include different types, e.g., simple features that directly look up embeddings and complex features that require complicated computations for online embeddings. Each type can be divided into either user/query features or item correlated features which include item features and interactive features between item and user/query. The user/query features are computed only once for online inference regardless of the number of candidate advertisements and require less computational consumption, while the item correlated features should be computed as many times as the number of advertisements to be scored, thus consuming much more computational resources. 
The $o_j$ values are determined by the above feature types and the detailed configurations are listed in Table~\ref{tab:priori_probabilities}. For other hyperparameters, $\gamma_1 = 1$, $\gamma_2 = 10^{-2}$, and $\gamma_3 = 10^{-7}$ are configured.

\begin{table}[]\footnotesize
\begin{tabular}{ccc}
\hline
\begin{tabular}[c]{@{}c@{}}feature type\\ index\end{tabular} & feature type & \begin{tabular}[c]{@{}c@{}} $o_j$ \end{tabular} \\
\hline
I   & \begin{tabular}[c]{@{}c@{}}Simple query/user features that\\ directly look up embeddings\end{tabular}      & 0.4 \\
II  & \begin{tabular}[c]{@{}c@{}}Simple item correlated features that\\ require complicated computations\end{tabular}            & 1.5 \\
III & \begin{tabular}[c]{@{}c@{}}Complex query/user features that\\ directly look up embeddings\end{tabular} & 1.0 \\
IV  & \begin{tabular}[c]{@{}c@{}}Complex item correlated features that\\ require complicated computations\end{tabular}       & 3.0 \\
\hline
\end{tabular}
\caption{Different types of features for the entire feature set with the values of $o_j$.}
\vspace{-7ex}
\label{tab:priori_probabilities}
\end{table}

\vspace{-2ex}
\subsection{Analysis of FSCD}\label{feature_selection_result}
\vspace{-0ex}
After the proposed FSCD process, the feature fields are ranked using the output parameter $\pmb{\delta}$. Table~\ref{tab:feature_distributions} lists the minimum, median, and maximum ranking indices for different types of features using Dropout FR~\cite{2017Dropout} and deep feature selection (DFS) method~\cite{2015Deep}. For the proposed FSCD in this paper, the ranking indices for the simple features are much smaller than those for the complex feature types (see median and maximum). The simple features tend to have a more forward ranking, while the complicated features are backward, which emphasizes the efficiency of the features. However, the minimum ranking index of type IV features is only $8$, which means that a feature with large complexity can still rank forward if it is truly effective for the training task. For the conventional methods, the rankings of the features are less dependent on the feature type. 

\begin{table}[]\footnotesize
\begin{tabular}{ccccl}
\hline
\begin{tabular}[c]{@{}c@{}}feature type\\ index\end{tabular} &
  \begin{tabular}[c]{@{}c@{}}number of \\ feature fields\end{tabular} &
  \begin{tabular}[c]{@{}c@{}}minimum\\ ranking\end{tabular} &
  \begin{tabular}[c]{@{}c@{}}median\\ ranking\end{tabular} &
  \multicolumn{1}{c}{\begin{tabular}[c]{@{}c@{}}maximum\\ ranking\end{tabular}} \\
 \hline
I   & 141 & 1/2/1 & 129/137/85 & 244/246/156 \\
II  & 5  &  15/31/18 & 36/44/42 & 99/52/159 \\
III & 68  & 3/1/4  & 137/131/185 & 246/243/231 \\
IV  & 32  & 7/3/8  & 108/58/229 & 178/239/246 \\
\hline
\end{tabular}
\vspace{1ex}
\caption{Feature ranking and distributions for different types of features, where the ranking results for different methods are provided and divided by slashes in order, including the Dropout FR method~\cite{2017Dropout}, the DFS method~\cite{2015Deep}, and the proposed FSCD.}
\vspace{-11ex}
\label{tab:feature_distributions}
\end{table}

The pre-ranking models with different feature field number $K$ are well trained based on Eq.~(\ref{L_s1}) with 2 billion samples. The area under the curve (AUC), latency, and CPU consumption for online inference of the different models are illustrated in Table~\ref{table:table_auc}. The results of the conventional feature selection methods~\cite{2015Deep,2017Dropout} are also provided for comparison. Specifically, when $K = 246$, the feature set of pre-ranking model becomes exactly the same as that of the ranking model. The results show that the proposed FSCD has slightly smaller AUC than the other methods when identical $K$ values are considered. However, the complexity is extremely reduced. When $K=100$, the AUC difference relative to that of the model for the other methods is only $0.0026$, while the complexity is approximately 30\% lower than that of the other methods for both latency and CPU cost. When $K > 100$, the AUC increases slowly, while the complexity increases significantly. Therefore, the pre-ranking utilizes the feature fields that have top-100 of $\delta$ values as the final feature set. 
\vspace{-2ex}
\begin{table}[]\footnotesize
\begin{tabular}{ccccc}
\hline
K                    & method        & AUC             & latency(ms)    & CPU(\%)         \\
\hline
\multirow{3}{*}{30}  & Dropout FR & 0.6910          & 4.32           & 13.26           \\
                     & DFS           & 0.6765          & 4.64           & 15.47           \\
                     & FSCD      & 0.6903(-0.007)  & 2.41(-44.2\%)  & 9.31(-29.7\%)  \\
 \hline
\multirow{3}{*}{50}  & Dropout FR & 0.6946          & 5.75           & 20.55           \\
                     & DFS           & 0.6788          & 5.19           & 20.57           \\
                     & FSCD      & 0.6927(-0.0019) & 3.54(-31.7\%)  & 12.10(-41.1\%)  \\
\hline
\multirow{3}{*}{100} & Dropout FR & 0.6975          & 6.48           & 30.82           \\
                     & DFS           & 0.6962          & 6.34           & 26.11           \\
                     & FSCD      & 0.6949(-0.0026) & 4.37(-31.0\%) & 19.18(-26.5\%) \\
\hline
\multirow{3}{*}{150} & Dropout FR & 0.6985          & 7.98           & 36.35           \\
                     & DFS           & 0.6972          & 7.85           & 35.41           \\
                     & FSCD      & 0.6958(-0.0027) & 4.74(-39.6\%) & 21.47(-39.3\%) \\
\hline
\multirow{3}{*}{200} & Dropout FR & 0.6983          & 8.10           & 37.26           \\
                     & DFS           & 0.6981          & 8.04           & 36.94           \\
                     & FSCD      & 0.6971(-0.0010) & 6.91(-14.0\%) & 30.09(-18.5\%) \\
\hline
whole (246)               & /             & 0.6990          & 9.02           & 40.60          \\
\hline
\end{tabular}
\vspace{1ex}
\caption{AUC results, latency, and CPU consumption for the first $K$ feature fields of the proposed FSCD and conventional methods. Each model is trained with 2 billion samples.}
\vspace{-8.5ex}
\label{table:table_auc}
\end{table}


\subsection{Performance Comparisons}\label{online_effects}

Table~\ref{tab:offline_effects} and Table~\ref{tab:online_effects} show the offline and online experimental results using the proposed effectiveness and efficiency based pre-ranking, while the original vector-product and COLD based pre-rankings are configured as benchmarks. Each model is trained with more than 200 billion real-world e-commerce samples from a log system to achieve the best online performance. Therefore, the offline AUCs are much larger than those shown in Table~\ref{table:table_auc}. The recall rate is defined as the preserved probability of pre-ranking model for the top-5 items ranked by the ranking model. The offline results in Table~\ref{tab:offline_effects} show that the proposed model is much more effective than the vector-product based model, while it is slightly less effective than COLD due to efficiency considerations. For the online effect, 30 continuous days of mobile real-world requests are evaluated and the results are shown in Table~\ref{tab:online_effects}. It shows that the proposed pre-ranking model obtains a good balance for online CTR and RPM performance and gains significant improvement in both CTR and RPM compared with its counterparts. 

\begin{table}[]
\begin{tabular}{ccc}
\hline
Methods   &  recall rate & offline AUC      \\
\hline
Vector-product based model     & 88\% & 0.695 \\
COLD model & 96\% & 0.738\\
The proposed model & 95\% & 0.737 \\
\hline
\end{tabular}
\vspace{0ex}
\caption{Offline results of the proposed pre-ranking model and the conventional models. Each model is trained with more than 200 billion samples.}
\label{tab:offline_effects}
\vspace{-7ex}
\end{table}

\begin{table}[]\footnotesize
\begin{tabular}{cccccc}
\hline
Methods    & \begin{tabular}[c]{@{}c@{}}input item\\ number\end{tabular}    & CTR     & RPM  & \begin{tabular}[c]{@{}c@{}}response\\ time\end{tabular} &  \begin{tabular}[c]{@{}c@{}}CPU\\ consumption\end{tabular}   \\
\hline
 \begin{tabular}[c]{@{}c@{}}vector-product\\ based model\end{tabular}       &    6000     & / & / &                58.4~ms                   &               79\% \\
COLD model  & 600 & +1.11\% & +1.04\% &                62.3~ms                &     85\% \\
The proposed model  & 800 & +1.54\% & +2.76\% &                59.9~ms                &     79\% \\
\hline
\end{tabular}
\vspace{0ex}
\caption{Online effectiveness and efficiency of the proposed pre-ranking model and the conventional models on real-world mobile online queries.}
\label{tab:online_effects}
\vspace{-10ex}
\end{table}

Finally, the efficiency is analyzed. Although a complicated pre-ranking with an interaction-focused architecture is adopted in the proposed approach, the feature selection method based on effectiveness and efficiency optimizes the computational complexity to a low degree. Moreover, the number of input item is reduced from $6000$ to $800$, which further reduces the overall complexity of the system. In this context, the online CPU consumption of the system is almost the same as that of the vector-product based model, while the response time is slightly increased. In regard to the COLD model, the response time and CPU consumption are both greater than those of the proposed approach, which means the proposed pre-ranking model is more efficient. The key metrics for efficiency are listed in Table~\ref{tab:online_effects} at a peak number of queries per second (QPS).

\vspace{-2ex}
\section{Conclusions}\label{sec-conclusion}

In this paper, to solve the problem of performance loss caused by the pre-ranking model with representation-focused architecture, a pre-ranking model based on feature selection with joint optimization for both effectiveness and efficiency is proposed in an interaction-focused architecture. The pre-ranking model with interaction-focused architecture, which is inherited from that of ranking model, utilizes the feature subset selected by the proposed learnable FSCD, which includes not only simple features but also interactive features with significant effectiveness. The experiments on offline training demonstrate the validity of the proposed FSCD method in both effectiveness and efficiency. Moreover, the offline and online effects in the real-world sponsored search system illustrate the performance improvements in the proposed pre-ranking model compared to the conventional benchmarks. The proposed pre-ranking has been utilized as an online running model for a real-world sponsored search system and has generated substantial revenue for the company.

\vspace{-2ex}
\appendix
\section{Derivation for Eq.~(\ref{equ:l_w_z}) and (\ref{alpha_theta}) }\label{appendix}

Assuming that $\pmb{v}$ and $\pmb{w}$ are subject to a joint Gaussian distribution~\cite{liu2017pbodl}, the joint distribution for $\pmb{v}$, $\pmb{w}$ and $\pmb{z}$ is
\vspace{-2ex}
\begin{equation}\label{priori}
P(\pmb{v}, \pmb{w}, \pmb{z}) =  \mathcal{N}(\pmb{v}, \pmb{w} | \pmb{0}, \Sigma) \prod_{j=1}^{M} \textup{Bern}(z_j | \theta_j),
\vspace{-2ex}
\end{equation}
where $\pmb{v}$, $\pmb{w}$ and $\pmb{z}$ are all learnable variables.

To optimize these variables, the loss function can be concluded by maximizing the posterior probability of $\pmb{v}$, $\pmb{w}$ and $\pmb{z}$ given the training samples $\mathcal{D}$. In this context, by applying Bayesian rule~\cite{rendle2012bpr},
maximizing $P(\pmb{v}, \pmb{w},\pmb{z}|\mathcal{D}) \propto P(\mathcal{D}|\pmb{v},\pmb{w},\pmb{z})P(\pmb{v},\pmb{w},\pmb{z})$ is equivalent to:
\begin{equation}
\begin{split}
\pmb{v}^*, \pmb{w}^*, \pmb{z}^* 
= \mathop{\arg\min}\limits_{\pmb{v}, \pmb{w}, \pmb{z}}-\textup{log}P(\mathcal{D}|\pmb{v}, \pmb{w},\pmb{z})-\textup{log}P(\pmb{v}, \pmb{w},\pmb{z}).
\end{split}
\end{equation}

The first part is the cross entropy, while the second part $-\textup{log}P(\pmb{v}, \pmb{w},\pmb{z})$ can be rewritten according to Eq.~(\ref{priori}) as
\vspace{-2ex}
\begin{equation}
\vspace{-2ex}
\begin{split}
-\textup{log}P(\pmb{v}, \pmb{w},\pmb{z}) = \lambda( ||w||^2 +  ||v||^2) +  \sum_{j=1}^{M} \alpha_j {z_j} + C,
\end{split}
\vspace{-3ex}
\end{equation}
where $\lambda$, $\alpha_j$ and $C$ are all constants derived by Eq.~(\ref{priori}). $ \lambda( ||w||^2 +  ||v||^2)$ is the common $l_2$ regularization term with weight $\lambda$. The nontrivial term $\sum_{j=1}^{M} \alpha_j {z_j}$ is a new regularization term derived by the Bernoulli distribution, where the regularization factor $\alpha_j$ is derived by Eq.~(\ref{Bern}) and (\ref{priori}) as
\begin{equation}\label{alpha_theta2}
\alpha_j = \textup{log}(1-\theta_j)- \textup{log}(\theta_j). 
\end{equation}
Then the total loss function can be written as Eq.~(\ref{equ:l_w_z}).

\newpage

\end{document}